# Experimental Study on Deuterium-Deuterium Thermonuclear Fusion with Interface Confinement

Darong Chen [1*#], Liang Jiang [2*†], Shuai Chen [3], Bao Wang [4], Dangguo Li [1], Peng Liang [1]

[1]*State Key Laboratory of Tribology, Tsinghua University, Beijing, 100084, China*
[2]*College of Automation, Wuxi University, Wuxi, 214105, China*
[3]*Materials Genome Institute, Shanghai University, Shanghai 200444, China*
[4]*Northeast Petroleum University，Daqing, 163318, China*

*\*Contributed Equally*
*[#]Email: chendr@mail.tsinghua.edu.cn*
*[†]Email: jiangliangthu@tsinghua.org.cn*

**Abstract:** Nuclear fusion is recognized as the energy of the future, and huge efforts and capitals have been put into the research of controlled nuclear fusion in the past decades. The most challenging thing for controlled nuclear fusion is to generate and keep a super high temperature. Here, a sonication system, combining with micro-scale fluid control techniques, was built to generate cavitation within a limited region. As bubbles being rapidly compressed, high temperature plasma generated interior leads to particle emissions, where a $Cs_2LiYCl_6$: $Ce^{3+}$ (CLYC) scintillator was used to collect the emission events. The pulse shape discrimination methods applied on captured signals revealed that only gamma ray events were observed in sonication with normal water as excepted, while obvious separation of neutron and gamma ray events was surprisingly identified in sonication with deuterated water. This result suggested that neutrons were emitted from the sonicated deuterated water, i.e. deuterium-deuterium thermonuclear fusion was initiated. This study provides an alternative and feasible approach to achieve controllable nuclear fusion and makes great sense for future researches on the application of fusion energy.

## 1. Introduction

As is known to all, the sun and stars are powered by the fusion of hydrogen into helium. If nuclear fusion can be successfully recreated on Earth, it holds out the potential of virtually unlimited supplies of low-carbon and low-radiation energy. In the past decades, plenty of efforts and huge amount of money has been put into fusion related researches, including inertial[1] and magnetic[2] confinement projects in worldwide, yet the way to practical application of fusion as energy supply is still long and unpromising. In 1996, Moss et al.[3] numerically studied the temperature inside a sonoluminescent cavitation bubble, and suggested that high temperature and pressure formed inside a collapsing bubble may satisfy the occurrence criteria of deuterium-deuterium thermonuclear fusion reaction, if the oscillating bubble is applied with proper acoustic activation. From then on, generating high temperature inside the cavitation bubble was considered to a promising alternative to achieve fusion in the lab[4]. Ever since the original work of single bubble sonoluminescence[5], huge efforts have been put on the understanding of oscillating cavitation bubble and its accompanying phenomena [6-14], and researchers found that the collapsing cavitation bubbles could be extremely compressed, resulting in high temperature and pressure[15-20]. In 2002, Taleyarkhan and his co-workers claimed that they observed neutrons emitted from imploding cavitation bubbles in a series of experiments[21-23], but their results have been greeted with criticisms ever since. Shapira et.al[24] and Barbaglia et. al[25] tried to reproduce the work, but both of them ended up unsuccessfully. This misfortune greatly struck the following research on related fields thereafter. However, the great potential applications of the controlled fusion initiated by cavitation bubbles still attract many researchers[26, 27].

The Astrophysical studies reported that both the gravity induced by the star matter itself and the electron/neutron degenerate pressure are crucial for the evolution of star[28]. As the gravity overcomes the effect of electron/neutron degeneracy pressure, the star would enter the gravitational collapse stage, which greatly increases the temperature inside the star[29]. These Astrophysical discoveries inspired researchers to explore the extreme pressure and temperature conditions in laboratory environment. Previous studies have proven that cavitation bubbles could be intensively compressed in a short time, and the material inside the bubbles would become plasma with high temperature[13, 30]. Just like the evolution of the stars, the matters inside the bubble, as well as the physical process these high temperature plasma experiencing, plays an important role in the collapse of the bubbles. The sheath formed near the bubble interface helps to constrain the energy state of the high temperature plasma. If the maters inside the bubble could be remarkably increased and the bubble could be compressed with high velocity, the temperature inside the bubble might reach to the criteria for thermal nuclear reaction. Hence, it makes sense to develop techniques that could bring the bubble into fiercer collapse with more matters inside.

During the oscillation of cavitation bubble, the movement of bubble wall of the collapse stage is surely faster than that of the expansion stage, and the mass transfer from liquid to vapor during the expansion stage plays an important role in the growth of the cavitation bubble. Accordingly, the greater this bubble wall acceleration difference between collapse stage and expansion stage exists, the larger the mass inside the bubble would be. In this paper, a unique sonication system was built to generate cavitation within a limited region, where a static electrical field was also imposed. Here, due to propagation and reflection of the ultrasonic wave within the confined space, the cavitation bubbles experienced much more complex driving pressure than ever. Thus, the mass transfer behavior through bubble interface into the bubble inside was considerably enhanced because of the fiercer bubble wall dynamics. Moreover, an independent neutron detection system, based on the CLYC scintillator, was established to study the particle emissions from collapsing cavitation bubbles. The pulse shape analysis (PSA)[31, 32] and the rise time analysis (RTA)[33, 34], applied on waveforms collected with the detector during the sonication experiments, both displayed obvious separation of neutron and gamma ray events, indicating the occurrence of deuterium-deuterium thermal fusion in sonication with deuterated water.

## 2. Experimental Setup

The arrangement of the sonication platform and the neutron detection system is shown in Fig. 1(a). A vibration horn with frequency of 15560 ±15 Hz was used to initiated cavitation in the sonication chamber, which was mounted on an electrical elevator. Fig. 1(b) shows the schematics of the sonication chamber. A coolant jacket with inlet temperature of 290 K was installed to cool the temperature of the solution during sonication. Tantalum pentoxide has a wide electrochemical window due to its high dielectric constant, so the electrolytic water reaction is hard to be initiated even when a high voltage was applied on. In this work, the voltage applied between the tip of the vibration horn and the bottom sample, covered with tantalum pentoxide, was set to be 15V to impose a static electrical field on the bubbles. The cavitation bubble dispersing height[35] was set to be 50μm, which is greater than the amplitude of the horn.

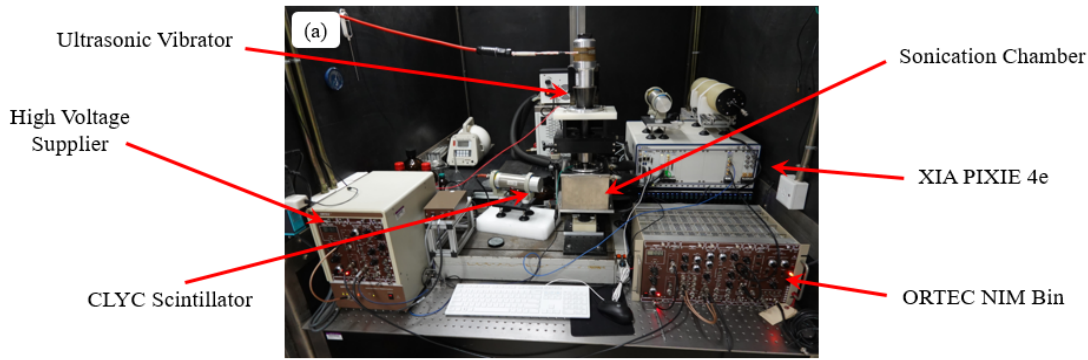

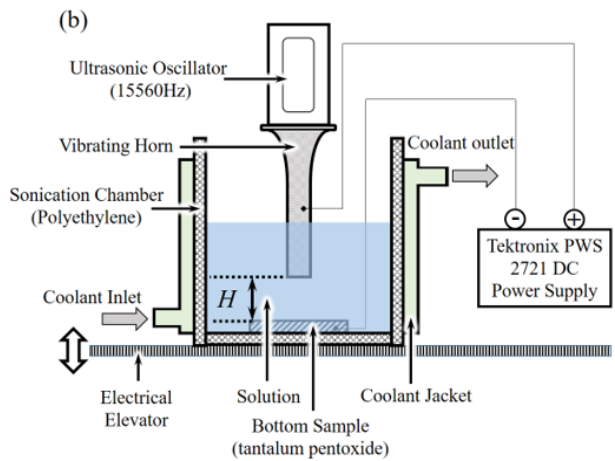
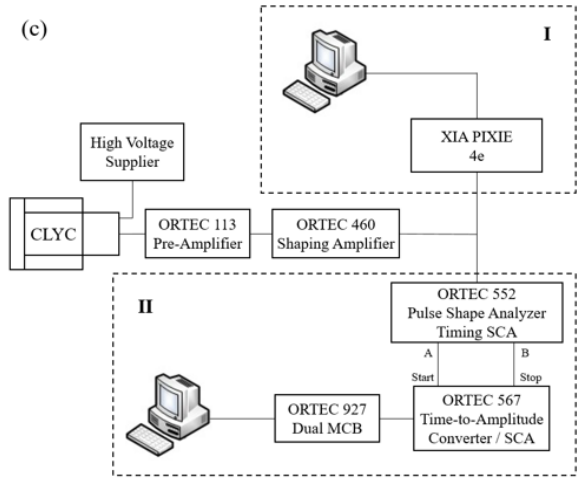

Figure 1. Experimental Setup (a). Arrangement of sonication platform and neutron detection system. (b). Schematics of sonication chamber. (c). Logic circuit diagram of neutron detection system.

The logic circuit diagram of the neutron detection system based on CLYC scintillator is presented in Fig. 1(c). The detection system contained one CLYC scintillator cell, optically coupled to a Hamamatsu R6231-100 photomultiplier (PMT), and an independent uninterrupted power supply (UPS) was utilized to reduce the electrical interference from the power net. After amplified, one output from the PMT was fed into the XIA PIXIE 4e module, a firmware incorporated with PSA algorithm. Another output was send to the RTA logic circuit diagram. In this way, the pulse shape discrimination was applied on the signals captured by the CLYC scintillator.

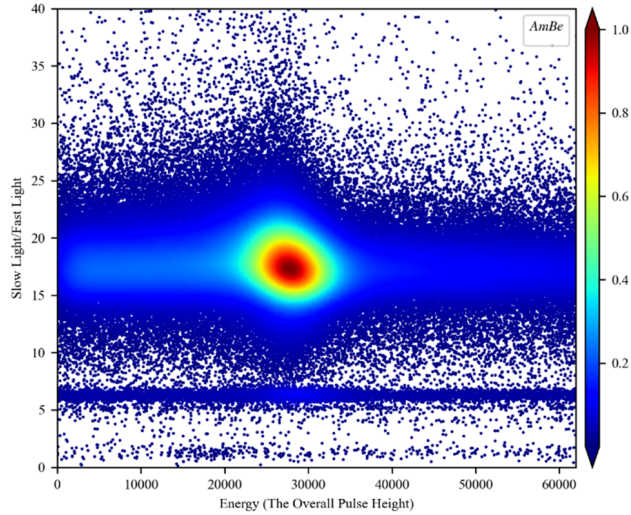

Figure 2. Scatter density plot of the signals captured by the detection system exposed to Am-Be Neutron source.

Before applied in the controlled experiments, the detection system was exposed to an Am-Be neutron source, and the emitted signals was collected and analyzed with the XIA PIXIE 4e module. The pulse shape discrimination (PSD) variable 'slow light /fast light', defined by the ratio of the integrated light in the tail of the PMT signal generated by an event in the scintillator to the integrated light around the signal's peak[36, 37], was employed to discriminate the neutrons from the gamma particles. Here, the PSD spectrum obtained with XIA PIXIE 4e module was presented in Figure 2. From the scatter density plot, it could be found that the signals are clearly divided into two regions, i.e. the upper branch with PSD variable larger than 10 and the lower branch with PSD variable smaller than 10. Given that the proton recoils induced by neutron have longer scintillation decay, the hits in upper branch represent the neutron events and the hits in the lower branch are considered as gamma ray events. For neutron events, most of their PSD variable mainly locates between 15 and 20, and the energy of these hits exhibits a Gaussian distribution with the peak around $2.8 \times 10^4$.

## 3. Results & Discussion

With the sonication system and the neutron detection system introduced above, a series of controlled experiments were carefully conducted in normal water and deuterated water, respectively. The volume of the sonication liquid was set to be 200ml in all experiments, and the duration of each sonication lasted 120s. The PSA and RTA results were presented in Fig. 3 and Fig. 4, respectively.

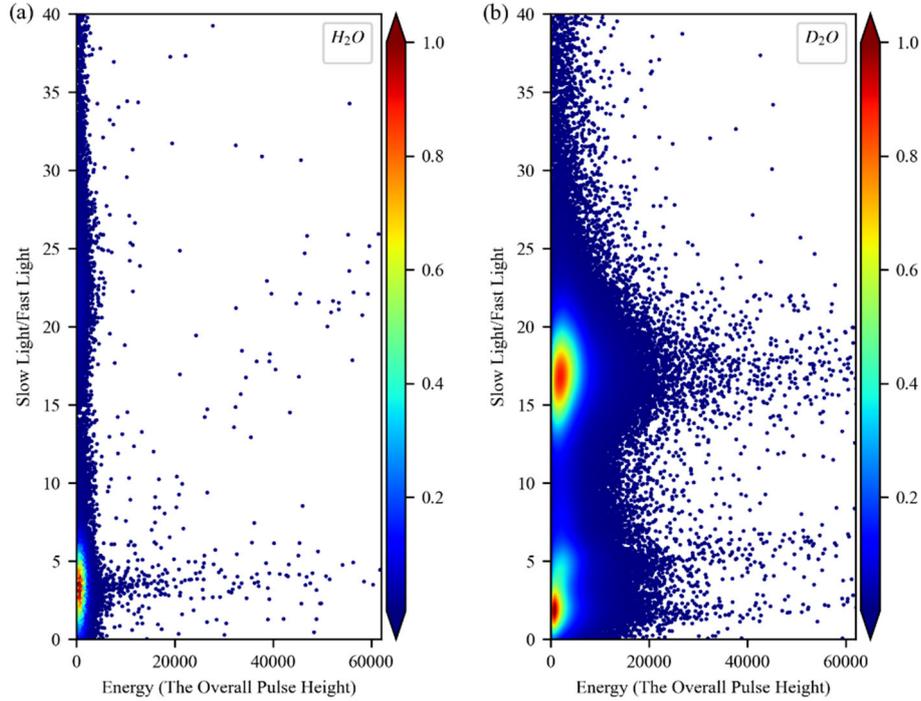

Figure 3. Pulse shape analysis of the signals. (a). Sonication with normal water (b). Sonication with deuterated water

In Fig. 3, the PSD spectra, for sonication in normal water and deuterated water respectively, accomplished with the algorithm embedded in XIA's Pixie spectrometer were presented. In Fig. 3(a), it could be noted that most of the signals have energy smaller than $2.0 \times 10^4$ on the chart, and the count of signals with PSD value ranging from 0 to 5 is obviously the highest. Otherwise, only a few signals with energy larger than $2 \times 10^4$ and PSD value greater than 10 are found and these signals just randomly scattered on the chart. However, in Fig. 3(b), when the sonication was induced in deuterated water, the count of signals with PSD value smaller than 10 increases drastically, and the energy of these lower PSD value signals also rise, mostly up to $2 \times 10^4$. Importantly, signals with PSD value ranging from 15 to 20 and energy greater than $2 \times 10^4$ are found to be densely scattering in Fig. 3(b), and this distribution pattern is in accordance with the neutron events shown in Fig. 2. This difference between Fig. 3(a) and Fig. 3(b) indicated that **neutron events was captured by the detection system from the sonicated deuterated water.**

When the sonication was performed in normal water, the collapse of the cavitation bubble would generate high temperature and high pressure inside the bubble, and then induce the transitions of electrons between different energy levels of the atoms, in which gamma ray particles could be emitted. Nevertheless, as in deuterated water, the extreme high physical condition inside the bubbles, resulting from the collapse of cavitation bubbles, could reach the level of deuterium-deuterium thermonuclear fusion, which emit considerable amount of neutrons and gamma ray particles. Accordingly, the detected spectrum in Fig. 3(b) presented two major concentrated regions, respectively corresponding to neutron events and gamma ray events, while only one major concentrated region was found in Fig. 3(a).

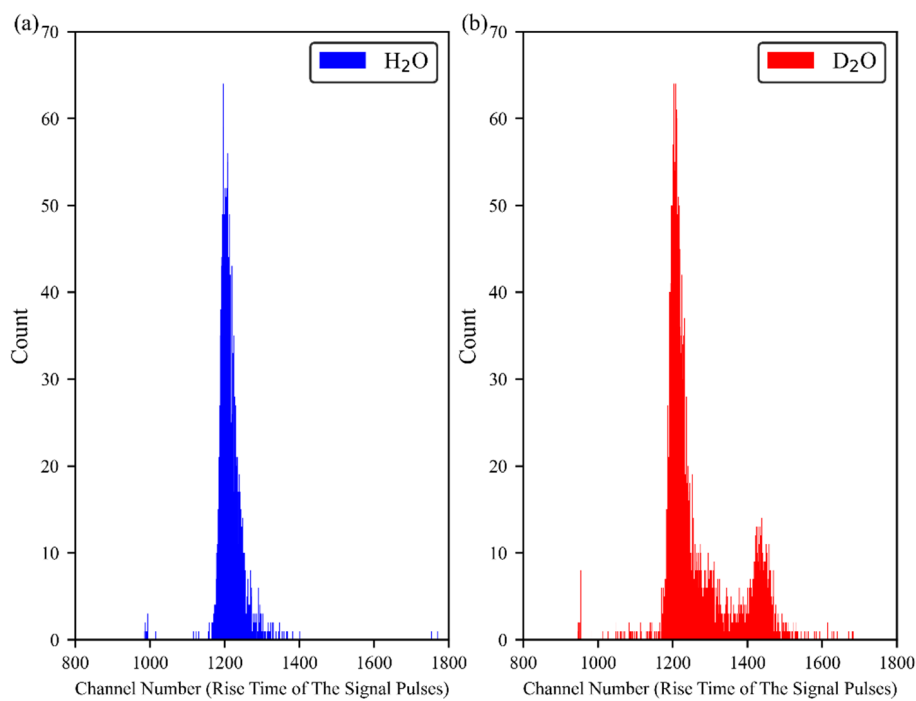

Figure 4. Rise time analysis of the signals. (a). Sonication with normal water (b). Sonication with deuterated water

Using the information of the rising time of the scintillation light pulse, Fig. 4 displayed the RTA results from sonication with normal water and deuterated water, respectively. In Fig. 4(a), only one single peak around 1200 was observed, which indicated that only one kind of particles were captured by the scintillator. In contrast, aside from the same peak as located in Fig. 4(a), another peak on larger channels, between 1400 and 1500, was found in Fig. 4(b), and obvious separation of these two peaks could be identified here. Since the proton recoils induced by neutrons have longer scintillation decay, the rising time of signals induced by neutron events ought to have larger value than that of gamma ray events. Thus, the two peaks in Fig. 4(b) undoubtedly correspond to the gamma ray events and the neutron events, respectively. Accordingly, the comparison of the RTA results in Fig. 4 also confirmed that **neutrons were** **detected when the sonication was generated in deuterated water**, which provided positive evidence of the occurrence of the deuterium-deuterium thermonuclear fusion in this work.

## 4. Conclusion

In this paper, intense cavitation, induced by ultrasonic vibrating horn within a limited region, was performed respectively in normal water and deuterated water, and the particle emissions behavior caused by collapsing cavitation bubbles was investigated with a detection system based on CLYC scintillator. The pulse shape discrimination procedures applied on the collected signals, carried out with pulse shape analysis and rise time analysis, implied that only gamma ray events were captured in sonication with normal water, while obvious separation of neutron and gamma ray events was identified in sonication with deuterated water. The detection accomplished here indicated that the deuterium-deuterium fusion was achieved, when the sonication was conducted in deuterated water. This study proposes a feasible approach to achieve controllable nuclear fusion and provides useful guidelines for future application of fusion energy.

## ACKNOWLEDGEMENTS

**This work was supported by the National Natural Science Foundation of China (Grant No. 50721004). Beside, we wish to thank for the financial support from Beijing Wind Capital Management Limited.**

# Supplementary Materials for "Experimental Study on Deuterium-Deuterium Thermonuclear Fusion with Interface Confinement"

Neutrons have mass but no electrical charge, so they cannot be directly detected as the ionized particles do. In practice, nuclear reactions between neutron and light nucleus, as well as the photon recoil induced by neutron in scintillating media, is widely used to detect neutrons.

In this supplement, pulse shape analysis was first discussed, and the method that used to discriminate neutron from gamma in the study was introduced in detail. Then, an extra neutron detection system that based on proportional counter, filled with $^3$He gas to 3040 Torr, was built to double check the neutron emission behavior in the study.

**Pulse Shape Discrimination**

In scintillation materials (e.g., BC-501A, EJ331, $Cs_2LiYCl_6$:Ce), ionizing particles produce light pulses classified by pulse shape discrimination as either proton-like or electron-like[1]. A proton-like pulse has relatively more long-lived scintillation components than an electronlike pulse, giving it an exaggerated tail. The vast majority of our cosmic background, due to muons and gammas, is electron-like. Neutrons, detected indirectly through proton recoil, are proton-like.

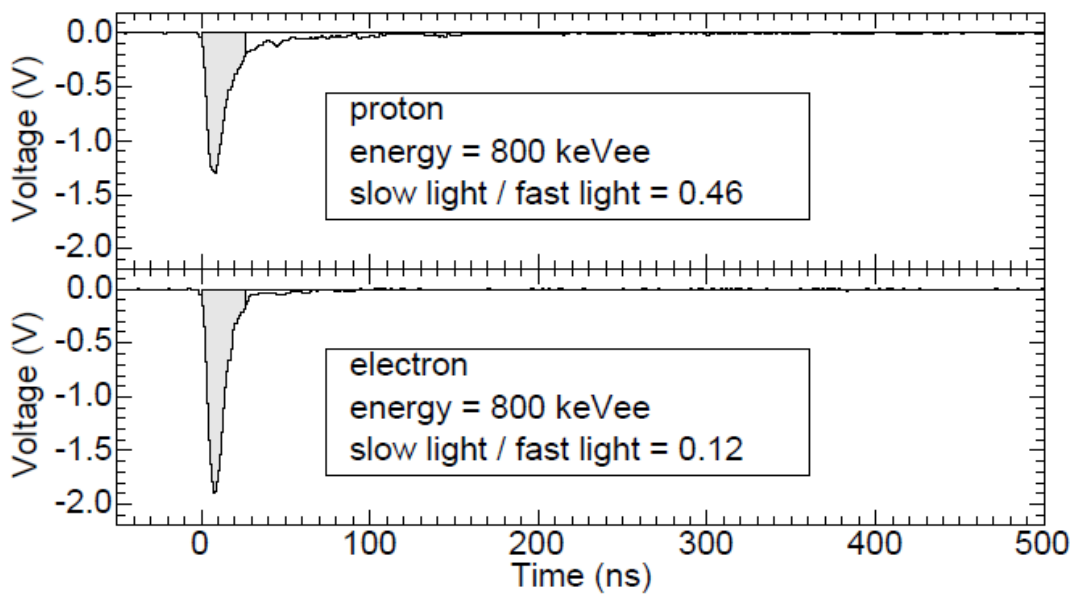

Figure 1. Typical proton-like and electron-like waveforms, both which have the same electron

equivalent energy 800 keVee[2].

Typical proton-like and electron-like waveforms collected during the nuclear fusion experiments[2] are shown in Figure 1. In some radiation detectors, analysis of pulse waveforms can provide additional information besides commonly obtained MCA pulse height spectra. Neutron/gamma pulse shape discrimination (PSD) is one form of such pulse shape analysis (PSA). It employs the observation that in certain scintillation materials light pulses display different decay characteristics depending on whether the energy deposited originates from a neutron or photon. Generally, one could capture all the neutron and gamma events with a high sampling speed digitizer and store the original waveform data into the hard disk, and use the procedure introduced by Dr. B. Naranjo[2] to discriminate the neutron hits from the gamma hits. However, capturing waveforms and applying offline PSA becomes increasingly cumbersome and time-consuming as event rates increase. As an alterative, XIA's Pixie-4e module was utilized to accomplish the PSA procedures in this study. This module incorporates the PSA algorithms into firmware within the data acquisition system.

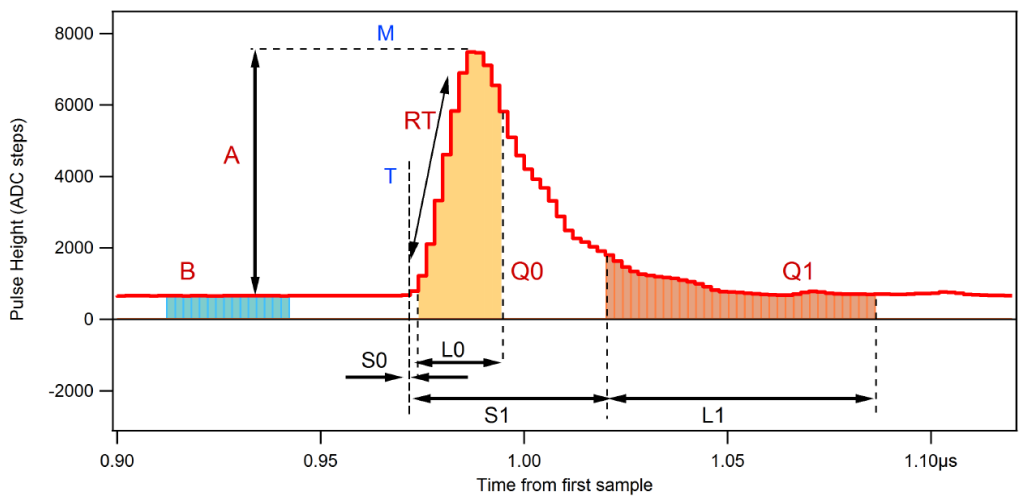

Fig.2: PSA parameters and results in XIA's Pixie-4e module

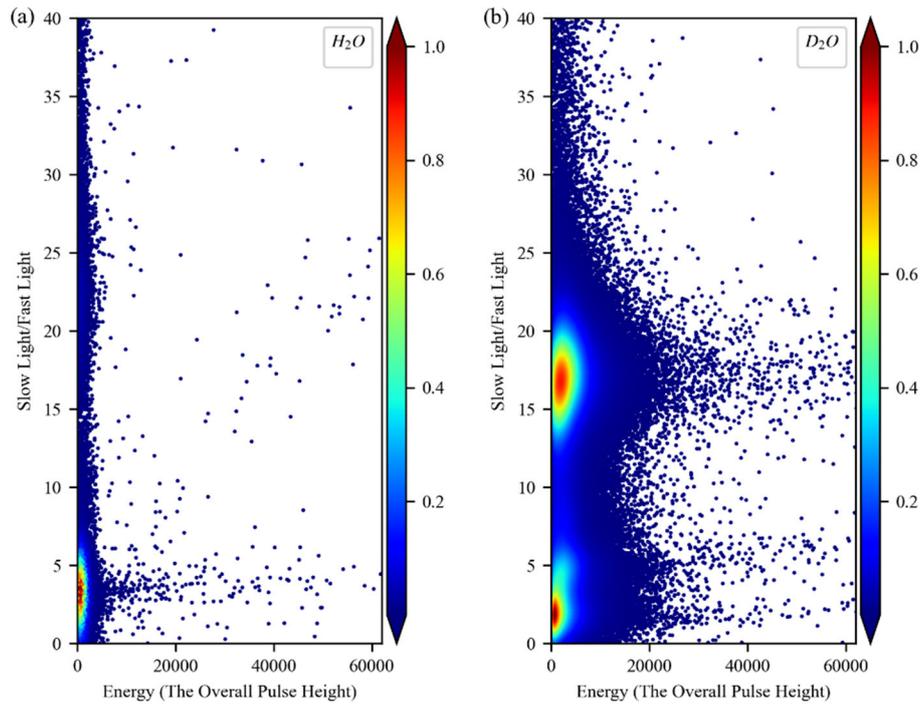

Figure 3. Pulse shape analysis of the signals collected with CLYC. (a). Signals captured in sonication with normal water (b). Signals captured in sonication with deuterated water

The PSA algorithms implemented inside the Pixie spectrometers are a version of the digital charge comparison method, in which two sums are accumulated over characteristic regions of a scintillator pulse[3]. Figure 2 defines four input values (black) that are specified by the user prior to the start of a run. In Fig.2, the DSP parameters S0 and S1 refer to the starting point of the sums referenced to the trigger T, and L0 and L1 are their lengths. Fig. 2 also shows the output values (red): PSA sums Q0 and Q1, a baseline sum B, and the amplitude A. Here in this study, S0 and S1 was set to be 0 and 32, L0 and L1 was set to be 16 and 64, respectively. The value 'slow light/fast light' introduced in this study could be calculated with the ratio between Q1 and Q0, i.e. Q1/Q0. In Figure 3, the signals collected with crystal scintillator, i.e. $Cs_2LiYCl_6: Ce^{3+}$ (CLYC), were analyzed and presented. As shown in Fig.3 (a), when the sonication was performed in normal water, only the signals with low level value of 'slow light/fast light' were found. On the other hand, clear separation of signals cloud be observed with deuterated water in Fig. 3 (b), as the sonication was carried out in deuterated water. Most importantly, this distribution pattern was in accordance with the calibrated spectrum as the detector exposed to the Am-Be neutron source.

## Helium-3 based neutron detection system

Here, a proportional counter encased with hydrogen-rich material, filled with $^3$He gas to 3040 Torr, was used to build a neutron detection system, to detect the possible neutrons emitted from the sonication in deuterated water.

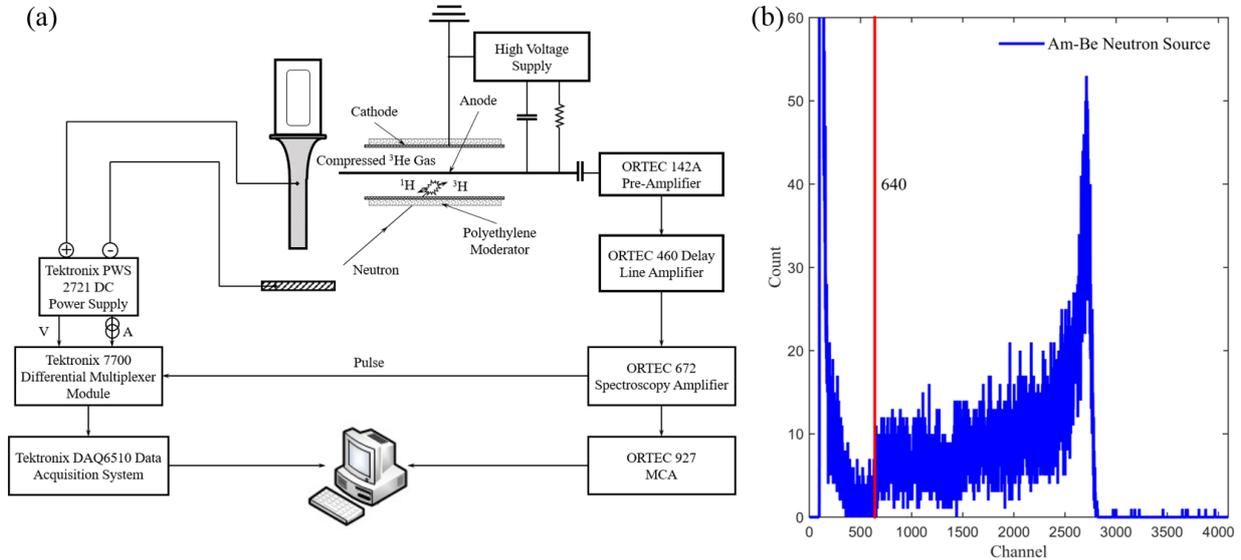

Figure 4. Helium-3 proportional counter based neutron detection system (a). Schematics of the $^3$He neutron detection system (b). Detected differential pulse height spectrum when exposed to an Am-Be neutron source.

Figure 4 showed the setup of the Helium-3 based neutron detection system. Outside the sonication chamber, a $^3$He proportional counter (model 253, LND) with diameter of 2 inches and length of 11.5 inches was mounted inside a polyethylene cylinder case. The wall thickness of the polyethylene cylinder case was designed to be 25 mm, so that the fast neutrons could be properly moderated to become slow thermal neutrons when they entered the counter, polarized by 1100 V. The output signal from the counter was fed into a pre-amplifier (model 142A, ORTEC), and then sent to a delay line amplifier (model 460, ORTEC) and a spectroscopy amplifier (model 672, ORTEC), sequentially. Here, the signal coming out of the spectroscopy amplifier was divided into two signal processing circuits. One of the circuits picked the signal and sent it into a multi-channel analyzer (model 927, ORTEC), to give the histogram of the detected signals. In another circuit, the emergence of the signal in the counter was captured by a differential multiplexer module (Tektronix 7700), which also recorded the variation of the voltage and current applied between the vibration horn and the bottom tantalum pentoxide piece. Eventually, the time stamp of these signals were recorded through the data acquisition module (Tektronix DAQ6510) into a computer.

When the $^3$He neutron detection system was exposed to an Am-Be neutron source, the pulse height histogram of the detected signals was obtained and shown in Fig. 4(b). As depicted in the histogram, a typical distribution with wall effects was observed, and this result soundly validated the effectiveness of the established $^3$He neutron detection system. From the histogram, it could be found that the wall effect of the counter started at channel 640, which marked the threshold channel to discriminate the neutron counts from the gamma ray interference.

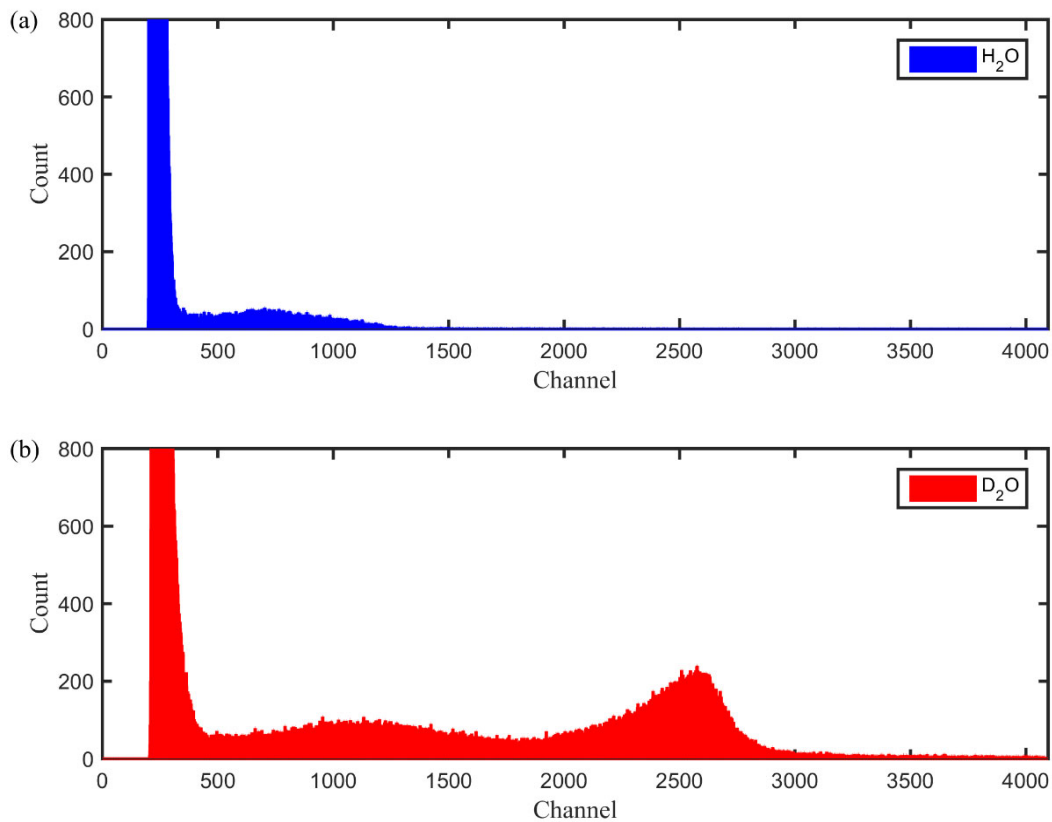

Figure 5. Comparison of spectrum given by $^3$He neutron detection system in Control Experiments with Normal Water and Deuterated Water. (a), Detected differential pulse height spectrum in Normal Water. (b), Detected differential pulse height spectrum in Deuterated Water.

The $^3$He neutron detection system was used to study the particle emissions phenomena in sonication with normal water and deuterated water, respectively. The collected spectra of the detected signals were shown in Fig. 5. When the sonication was induced in normal water, the signals captured by the detector only distributed on low level channels, and no obvious peak was found on channels exceeding 640, as shown in Fig. 5(a). However, as depicted in Fig. 5(b), the differential pulse height spectrum here presented a similar profile as that in Fig. 4(b), and the unique wall effect could be clearly observed in the spectrum. This difference between experiments in normal water and deuterated water implied that the deuterium-deuterium thermal fusion might be achieved in the sonication of deuterated water, which also confirmed the repeatability of our experimental results.

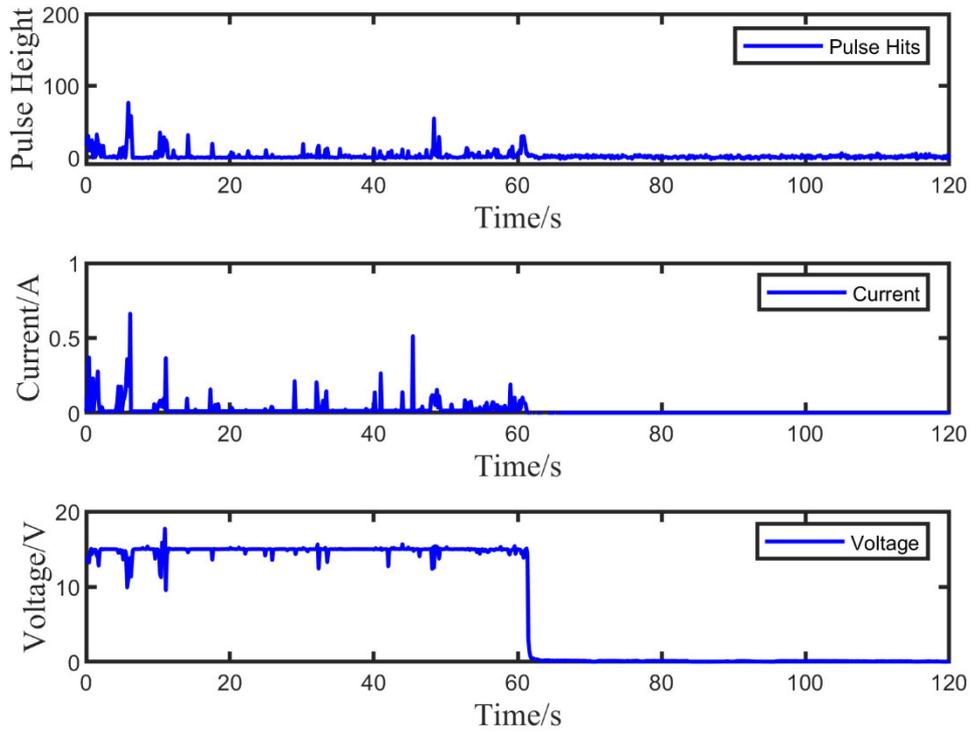

Figure 6. Variation of the pulse height output from the $^3$He proportional counter, the current and the voltage output from the DC power supply recorded in sonication with normal water. (a). The pulse height (b). The current (c). The voltage

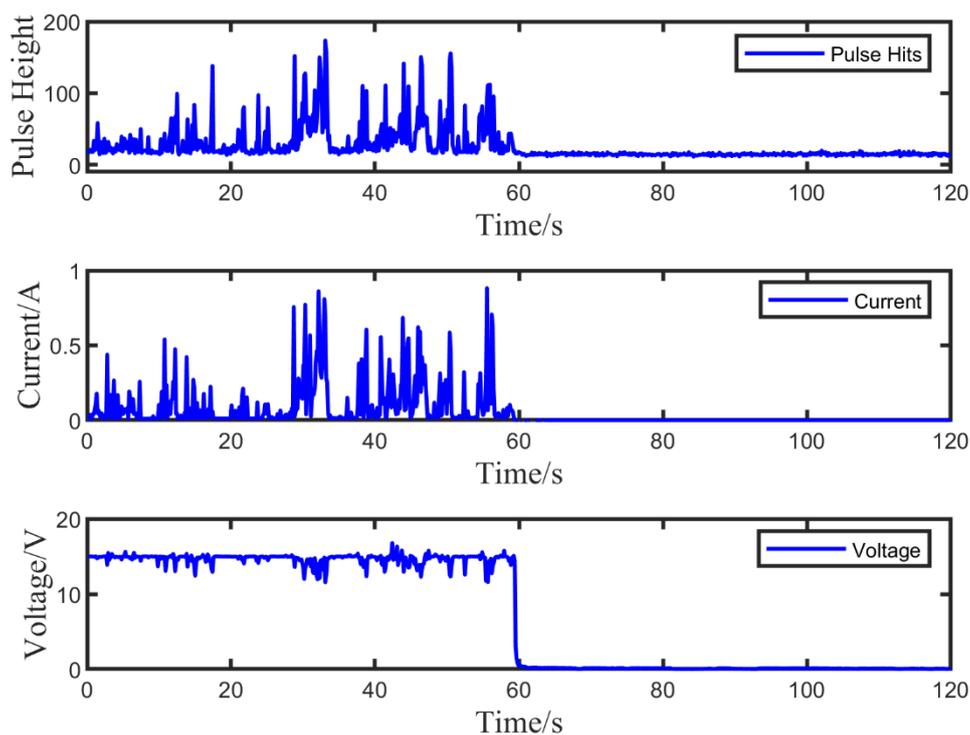

Figure 7. Variation of the pulse height output from the $^3$He proportional counter, the current and the voltage output from the DC power supply recorded in sonication with deuterated water. (a). The pulse height (b). The current (c). The voltage

To study the influence of the static electrical field brought by the DC power supply, detailed experiments with normal water deuterated water were designed and performed. During these experiments, the output voltage of the DC power supply was set to be 15V in the first half 60s, while it was set to be 0V in the second half 60s. The emergence of the signal captured by the $^3$He proportional counter, as well as the change of the output current and voltage from the DC power supply, was recorded and presented in Fig. 6 and Fig. 7, respectively for normal water and deuterated water. As shown in Fig. 6, when the voltage from the DC power supply was set to be 15V in the first half 60s, the $^3$He proportional counter captured a lot of signals from the reaction chamber, and current and the voltage also presented certain level fluctuation as the sonication went on. However, as the sonication was performed in deuterated water, the signals captured with the $^3$He proportional counter displayed a steep increase, as well as the current and the voltage recorded during the sonication. It is known that plasma could be generated inside the bubble when collapse. Under such high temperature as inside the bubble, the discharge of plasma could happen. This process of discharge of plasma would change the distribution of charge outside the plasma sheath, which eventually cause the generation of current. So that, it could be noticed that every time the counter captured a signal, the current and the voltage changed accordingly, i.e. the changes of these three variables were basically synchronous and displayed certain correlation. Nevertheless, as the voltage from the DC power supply decreased to 0V in the second half 60s, the fierce fluctuation of these variables recorded in both cases disappeared. Hence, it could be concluded that the static electrical field was crucial for the physical process during the sonication, especially the particle emissions behavior.